\documentclass[prl,aps,floats, superscriptaddress,longbibliography, twocolumn]{revtex4-1}
\usepackage{gensymb}
\usepackage{amsmath}
\usepackage{amssymb}
\usepackage{graphics}
\usepackage{graphicx}
\usepackage{hyperref}
\usepackage{bm}

\hypersetup{colorlinks,linkcolor=blue,urlcolor=blue,citecolor=blue}

\begin{document} 

\title{Flat bands and perfect metal in trilayer moir\'e graphene}

\author{Christophe Mora}
\affiliation{Laboratoire de Physique de l’Ecole normale sup\'erieure, ENS, Universit\'e PSL, CNRS, Sorbonne Universit\'e, Universit\'e Paris-Diderot, Sorbonne Paris Cit\'e, Paris, France}
\author{Nicolas Regnault}
\affiliation{Laboratoire de Physique de l’Ecole normale sup\'erieure, ENS, Universit\'e PSL, CNRS, Sorbonne Universit\'e, Universit\'e Paris-Diderot, Sorbonne Paris Cit\'e, Paris, France}
\author{B. Andrei Bernevig}
\affiliation{Department of Physics, Princeton University, Princeton, New Jersey 08544, USA}

\begin{abstract}
  We investigate the electronic structure of a twisted multilayer graphene system forming a moir\'e pattern. We consider small twist angles separating the  graphene sheets and develop a low-energy theory to describe the coupling of Dirac Bloch states close to the K point in each individual plane. Extending beyond the bilayer case, we show that,  when the ratio of the consecutive twist angles is rational, a periodicity emerges in quasimomentum space with moir\'e Bloch bands even when the system does not exhibit a crystalline lattice structure in real space. For a trilayer geometry, we find  flat bands in the spectrum at certain rotation angles. Performing a symmetry analysis of the band model for the trilayer, we prove that the system is a perfect metal in the sense that it is gapless at all energies. This striking result originates from the three Dirac cones which can only gap in pairs and produce bands with an infinite connectivity. The full gapless property is protected by an emergent particle-hole symmetry valid at sufficiently small angles.
\end{abstract}
  
\date{\today}

\maketitle

Two parallel layers of graphene twisted by a small angle exhibit a moir\'e pattern~\cite{pong2015}, with a lattice periodicity much larger than the graphene unit cell, and non-trivial electronic properties~\cite{Bistritzer2010}, such as band-flattening at certain {\it magic} angles~\cite{MacDonald_M-Model,laissardiere2010,sanjose2012,dossantos2012,TB_TBG,tarnopolsky2018}. Flat bands present a reduced kinetic energy, thereby artificially boosting electron correlations~\cite{Kim3364}. The recent discovery of correlated insulating phases at half-filling and possibly unconventional superconductivity~\cite{cao_TBG1,cao_TBG2,carr2018,yankowitz2018} have unveiled bilayer moir\'e graphene as a tunable device for exploring novel correlated states, at zero or finite magnetic field, spurring intense theoretical work in this direction~\cite{yuan2018,wu2018,padhi2018,po2018,isobe2018,dodaro2018,baskaran2018,huang2018,you2018,wux2018,roy2018,xu2018,zhangy2018,koshino2018,kang2018,ray2018,guinea2018,wu2018,lian2018,kang2018-2,biao2018}. On the other hand, moir\'e bands were also investigated for their topological properties~\cite{sanjose2013,gail2011,Po2018b,song2018,ahn2018,zou2018band,liu2018} and topological phase transitions were identified close to the magic angles~\cite{song2018,hejazi2018}.

In view of the great wealth of correlation and topological phenomena occuring with moir\'e bilayer graphene, it is desirable to extend studies to multilayer, and specifically trilayer geometries, in which moir\'e patterns also appear for small rotation angles. Flat bands in bilayer result from an interplay between the K (or K') Dirac points in each layer and the situation with three or more Dirac points has yet to be explored. At low energy and close to half-filling, the band structure is formed only from the electron states of the Dirac cones in each layer~\cite{TBGstr}. A moir\'e band theory describing such as state, that does not require a crystalline lattice, is built in Ref.~\cite{MacDonald_M-Model}. In this paper, we extend this theory to multilayer graphene and discuss in depth the symmetry and topology for three layers. We find magic angles of vanishing Dirac velocities; they are not related to a complete flattening of the spectrum, rather by a flattening along certain symmetry lines. We also characterize the different moir\'e bands by the irreducible representations they generate at the high-symmetry points and lines. Based on the compatibility between these representations~\cite{Bernevig_TQC,Vergniory2017}, we are able to prove the remarkable result that
all bands are connected such that no subset of bands can be energetically isolated from the others. The most obvious consequence is that the system remains metallic at arbitrary energy. This property requires particle-hole symmetry - neglecting the band curvature in the  vicinity of the original K points.

\noindent {\it Bloch band coupling.} We detail the derivation of the band structure of the moir{\'e} pattern in twisted multilayer graphene. When the twisting angle is small, a moir{\'e} pattern is formed by the interference of lattices between the different layers.
Restricting the analysis to Dirac fields near the K points of each layer~\cite{TBGstr,MacDonald_M-Model,TB_TBG} (see also Ref.~\cite{amorim2018}) and assuming a local short-range tunnel amplitude between atoms in consecutive planes, one derives the following Hamiltonian
\begin{equation}
\begin{split}
  H^{(ab)}(\delta \mathbf{p}_a, \delta \mathbf{p}_b)&  = v_{F}\delta\mathbf{p}\cdot\boldsymbol{\sigma} \delta_{a,b} \\ &+  w^{ab} \sum_{j=1}^{3}\delta_{\delta\mathbf{p}_a,\delta\mathbf{p}_b+\mathbf{q}_{j}^{a,b}}T^{j} \label{hamiltonianbetweenlayesrab}
\end{split}
\end{equation}
where $w^{ab}$ are hopping energies between the neighboring layers $a$ and $b$. The first term in Eq.~\eqref{hamiltonianbetweenlayesrab} represents the Dirac cones in each layer and $\delta\mathbf{p}$ is a small momentum deviation from the $K$ point for the layer $a$. We have introduced the matrices 
\begin{equation}
 T^{j+1} = \sigma_0 + \cos ( 2 \pi j/3) \sigma_x  + \sin ( 2 \pi j/3) \sigma_y
\end{equation}
associated with the three symmetric momentum directions $\mathbf{q}_{1}^{a,b}=M_{a b}\mathbf{K}-\mathbf{K}$, $\mathbf{q}_{2}^{a,b}=C_{3z}\mathbf{q}_{1}^{a,b}$, $\mathbf{q}_{3}^{a,b}=C_{3z}\mathbf{q}_{2}^{a,b}$. $M_{a b}$ the rotation around $z$ with the twist angle $\theta_{ab}$ separating the layers $a$ and $b$ and $C_{3z}$ with angle $2 \pi/3$. For small twist angles, $\mathbf{q}_{1}^{a,b}$ is perpendicular to ${\bf K}$ so that  all the ${\bf q}_j^{ab}$ are parallel for fixed $j$ (see below).

The magnitudes of the vectors $| \mathbf{q}_{j}^{a,b} | = 2 |\mathbf{K}| \sin(\theta_{ab}/2)$ depend on the twist angles. The second term in Eq.~\eqref{hamiltonianbetweenlayesrab} couples momenta $\delta\mathbf{p}_a$ and $\delta\mathbf{p}_b+\mathbf{q}_{j}^{a,b}$ in layers $a$ and $b$ which generates a lattice in momentum space for each pair of consecutive layers. To make the calculation tractable and maintain an emergent periodicity in momentum space regardless of whether the multilayer system itself is crystalline, one has to assume that the ratios of twist angles are rational numbers. This periodicity results in moir\'e bands which we compute numerically and classify according to their irreducible representations at the high-symmetry points and lines. On the contrary, if the twist angles were incommensurate, or the ${\bf q}_j^{ab}$ vectors are not parallel for fixed $j$, then successive applications of the Hamiltonian~\eqref{hamiltonianbetweenlayesrab} would reach arbitrary momentum and the moir\'e periodicity would be absent.

\begin{figure}
\begin{centering}
\includegraphics[width=0.8\linewidth]{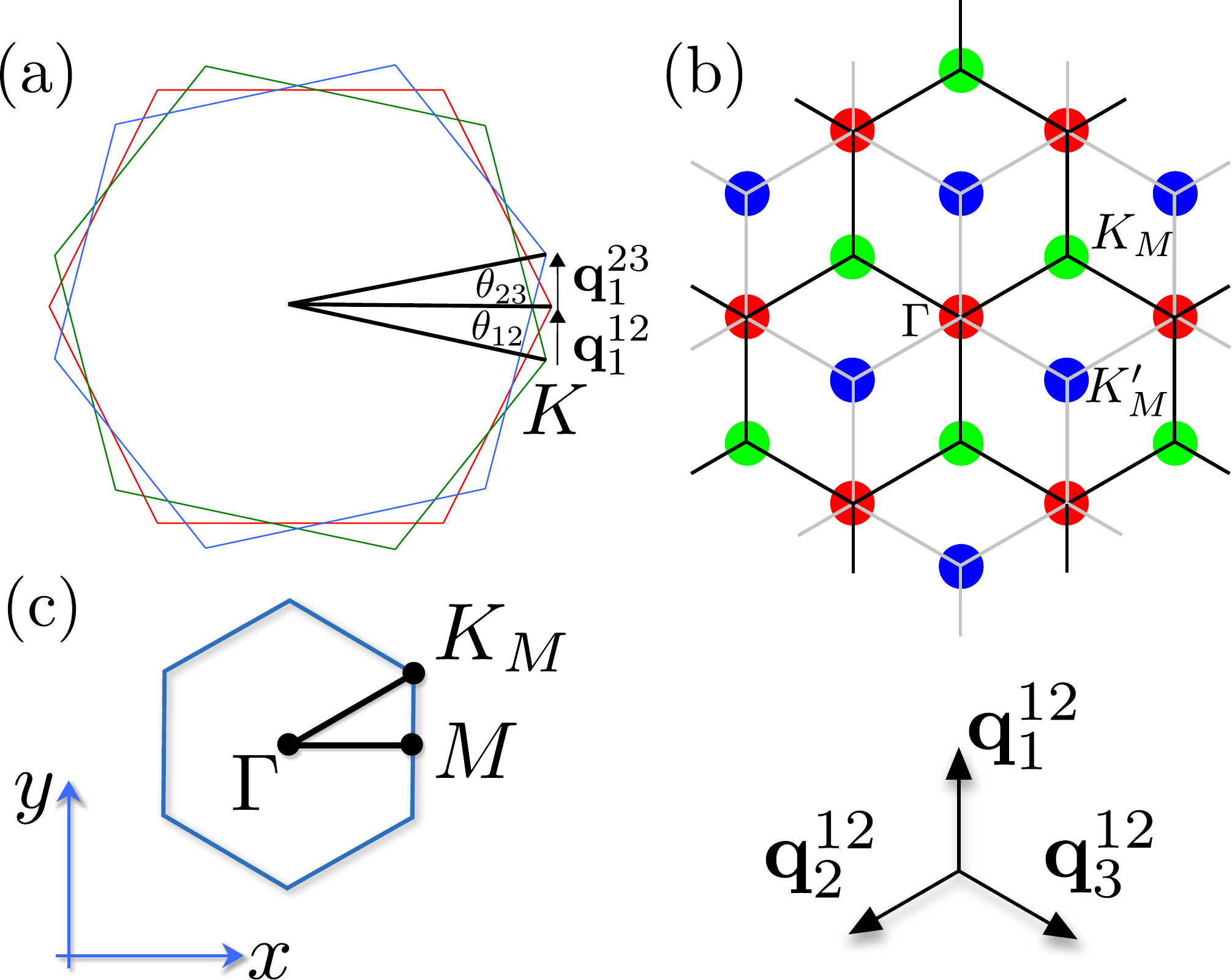}
\par\end{centering}
\caption{\label{fig1} Trilayer graphene with the same rotation angle $\theta_{12} = \theta_{23}$ between consecutive layers. (a) Original Brillouin zones in each layer with their respective K points. (b) k-space lattice generated by the vectors ${\bf q}_j^{ab}$ (for ${\bf q}_j^{12} = {\bf q}_j^{23}$). Green, red and blue sites belong respectivement to the layers $1,2,3$. (c) Moir\'e Brillouin zone with high-symmetry points $\Gamma$, $K_M$, $K_M'$, $M$ and high-symmetry lines (in black).}
\end{figure}

The Hamiltonian~\eqref{hamiltonianbetweenlayesrab} reduces to the model of Ref.~\cite{MacDonald_M-Model} for bilayer and we henceforth focus on the trilayer geometry. We take the rotation angle $\theta_{12}$
as a reference and introduce the moir\'e magnitude $k_D = 2 |\mathbf{K}| \sin(\theta_{12}/2)$. We rescale all momenta by $k_D$ and the Hamiltonian as $\tilde H = H/(v_F k_D)$. Fixing the direction of $\mathbf{q}_{1}^{a,b}$ along $y$, we use the complex notation
\begin{equation}
   q_1^{23}=  e^{i \frac{\pi}{2}};\;\;\;\;q_2^{23}= e^{i \frac{7\pi}{6}};\;\;\;\; q_3^{23}= e^{-i \frac{\pi}{6}};
\end{equation}
and $q_j^{12}= (p/q) q_j^{23}$ for all $j$, where $p$ and $q$ are coprime integers. The Hamiltonian~\eqref{hamiltonianbetweenlayesrab} can then be written as
\begin{equation}\label{hamiltonianbetweenlayers}
  \tilde H_{Q_m Q_n} (k) = ({\bf k}- {\bf Q}_m) \cdot {\bm \sigma} \delta_{mn} + \alpha \sum_j T^j \delta_{Q_m, Q_n - q_j^{mn}}
\end{equation}
where we assume a uniform tunnel amplitude $w_{ab} = w$ and introduce the dimensionless coupling $\alpha = w/(v_F k_D)$ between Dirac cones. The vectors ${\bf Q}_m$ form a $k$-space lattice, see Fig.~\ref{fig1}b, where each site is associated to a specific layer.

\begin{figure}
\begin{centering}
\includegraphics[width=1.\linewidth]{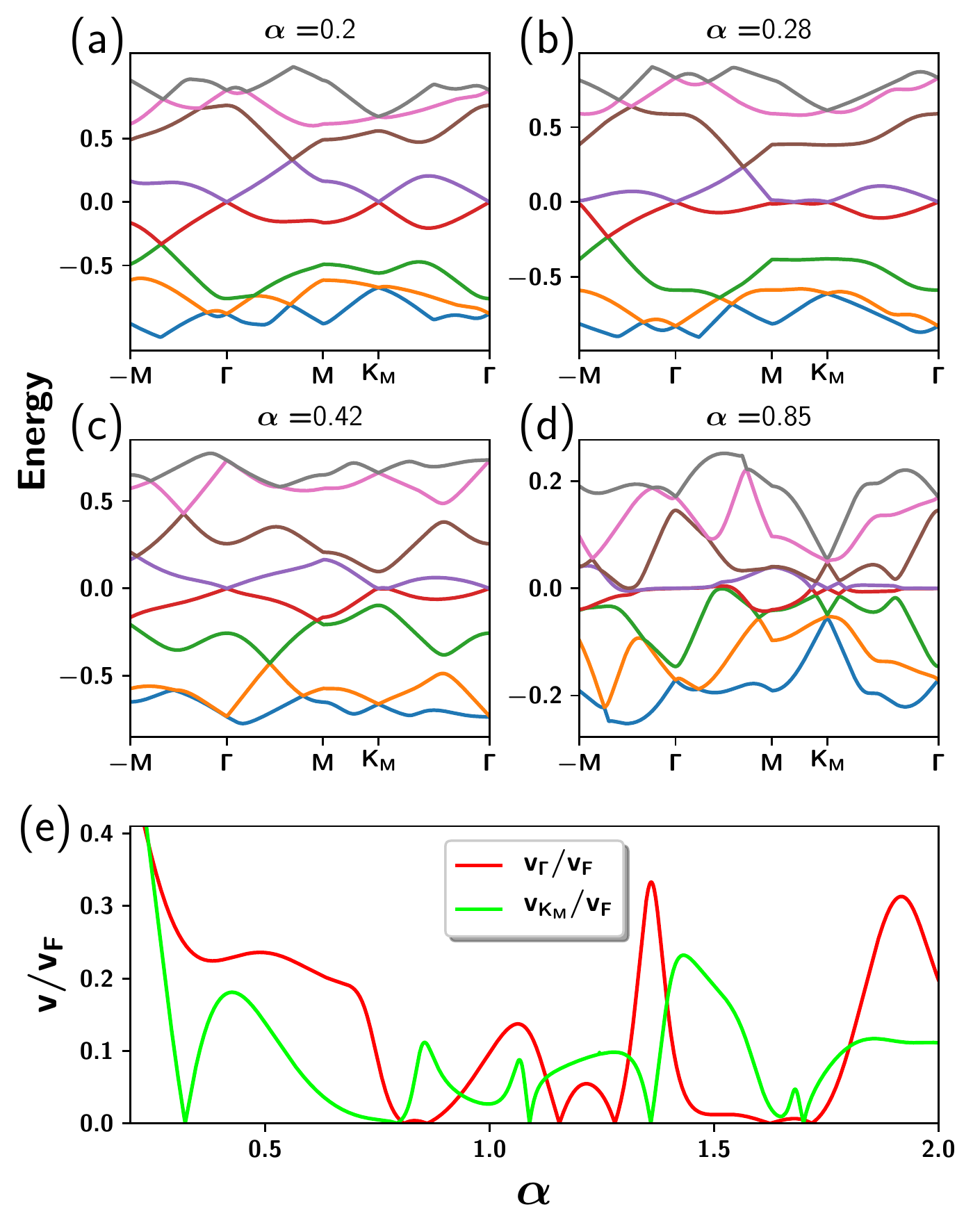}
\par\end{centering}
\caption{Moir\'e bands (a-d) and renormalized Dirac-point velocities (e) at the symmetric points $\Gamma$ (red) and $K_M$ (green).  The eight bands closest to zero energy are represented along the moir\'e Brillouin zone trajectory $-M \to \Gamma \to M \to K_M \to \Gamma$ for $\alpha = 0.2, 0.28, 0.42, 0.85$ (a-d). The particle-hole symmetry discussed in the main text - sending $k$ to $-k$ and $E$ to $-E$ is clearly visible along the path  $-M \to \Gamma \to M$. (e) Velocities of the two Dirac cones at $\Gamma$ and $K_M$ as function of $\alpha$.
  \label{fig2}}
\end{figure}

\noindent {\it Equal twist angles.} We first consider the most symmetric case of evenly rotated planes where $q_j^{12} = q_j^{23}$. A representative set of moir\'e spectra obtained from Eq.~\eqref{hamiltonianbetweenlayers} with different values of the coupling $\alpha$ is displayed in Fig.~\ref{fig2}a-d.
The moir\'e bands exhibit a rich structure. The first remarkable feature is that all bands are connected: it is impossible to isolate a set of bands which are detached from the rest.
We provide below a formal proof for this statement based on irreducible representations at symmetric points and lines, and afterwards extend it to arbitrary $p$ and $q$.

Three Dirac cones are attached at zero energy to the points $\Gamma$, $K_M$ and $K_M'$ (see Fig.~\ref{fig1}b) as $\alpha$ is varied. We display the corresponding Dirac velocities of the cones at $\Gamma$ and $K_M$ (with $K_M'$ velocity linked by symmetry to that of $K_M$) in Fig.~\ref{fig2}e and find a set of {\it magic angles} - in analogy with the bilayer case - where one of these velocities vanishes. The difference with the bilayer case is that these magic angles are not associated with a flattening of the whole spectrum which would be at odds with the fully connected band structure. However, we do see a flattening of part of the spectrum close to magic angles: on the $M-K_M$ line (for the second levels) when $\alpha = 0.28$, close to $\Gamma$ when $\alpha = 0.85$, for the first two magic angles. This flattening along one-dimensional directions in $k-$space open interesting perspectives for the realization of exotic correlated many-body physics.

\noindent {\it Symmetries} 
The moir\'e reciprocal lattice vectors
$b_1 = q_1^{12} - q_2^{12}$, $b_2 = q_1^{12} - q_3^{12}$
generate the whole lattice in Fig.~\ref{fig1}b. The different layers (colors) are coupled by the $q_j^{a,b}$ vectors. Bloch periodicity takes the form
\begin{equation}
  \tilde H(k-b_i) = V^{b_i} \tilde H(k) V^{b_i \dagger}, \quad V^{b_i}_{Q_m, Q_n} = \delta_{Q_n, Q_m+b_i}.
\end{equation}
The spectrum is thus invariant upon shifting the origin of $k$ by a combination of $b_1$ and $b_2$.

The moir\'e lattice also transforms into itself by the action of a $2 \pi/3$ rotation $C_{3z}$ around $\Gamma$.
In the Hamiltonian language, the corresponding operator is given by  $\mathcal{C}_{3z} =\exp(i 2\pi\sigma_z /3 ) \delta_{Q_m, C_{3z} Q_n}$. We note that if $Q_n$ is in a given layer so is $C_{3z} Q_n$, such that the three layers are not mixed by the rotation. The $C_{2x}$ symmetry operates a reflexion accross the x-axis going through the $\Gamma$ point (red in Fig.~\ref{fig1}b). As such, $C_{2x}$ sends lattice sites of layer $1$ to $3$ and viceversa but keeps layer $2$ invariant. The corresponding symmetry operator is $\mathcal{C}_{2x} =\sigma_x \delta_{Q_m, C_{2x} Q_n}$. The two symmetry operators induce the Hamiltonian transformation
\begin{equation}
    \mathcal{C}_{3z}   \tilde H(k) \mathcal{C}_{3z}^\dagger  =  \tilde H(C_{3z}k),  \quad
\mathcal{C}_{2x}  \tilde H(k) \mathcal{C}_{2x}^\dagger =  \tilde H(C_{2x}k),
\end{equation}
which leave the spectrum invariant. The antiunitary $C_{2z} T$ symmetry acts locally on the moir\'e lattice. It takes complex conjugation $K$ and reverses the pseudo-spin direction. It is represented by the operator $\mathcal{C}_{2z} \mathcal{T} =\sigma_x \delta_{Q_m, Q_n} K$, which squares to $1$ and commutes with the Hamiltonian $\tilde H(k)$. It also commutes with the spatial symmetries $\mathcal{C}_{3z}$ and $\mathcal{C}_{2x}$.

The moir\'e model also possesses a unitary particle-hole (p-h) symmetry. The original $k\cdot p$ Dirac Hamiltonian of the single layer graphene sheet $H(k) = \vec{k} \cdot \vec{\sigma}$ has a unitary particle-hole symmetry $H(k) = -H(-k)$ due to the absence of $k^2$ terms in the Hamiltonian. Since our model is based on this low-energy expansion, it retains a similar symmetry with the operator
\begin{equation}\label{phsym}
\mathcal{P} = \delta_{Q_m, -Q_n}\zeta_{Q_n}
\end{equation}
where $\zeta_{Q_n}$ is $+1$ for $Q_n$ belonging to the lower and top layers and $-1$ for the middle layer. With this, we have $\mathcal{P}^2 = \mathcal{P}\mathcal{P}^\dagger =1$.
Importantly, one checks that $\mathcal{P}$ commutes
with all other symmetry operators, $\mathcal{C}_{3z}$, $\mathcal{C}_{2x}$ and $\mathcal{C}_{2z} \mathcal{T}$, and satisfies
\begin{equation}\label{phsym2}
  \mathcal{P}   \tilde H(k) \mathcal{P}^\dagger = -  \tilde H(-k).
\end{equation}
This is to be contrasted with the p-h operator $\mathcal{P}_{\rm bi}$ identified~\cite{song2018} for moir\'e bilayer which has: (i) $\mathcal{P}_{\rm bi}^2=1$, and (ii) anticommutes with $\mathcal{C}_{2x}$, {\it i.e.} $\{ \mathcal{P}_{\rm bi}, \mathcal{C}_{2x} \} =0$, instead of the commutation found for $\mathcal{P}$.
\begin{table}
\begin{tabular}{lrrr|lrr|lrr|lrr}
\hline 
& $\Gamma_{1}$ & $\Gamma_{2}$ & $\Gamma_{3}$ &  & $M_1$ & $M_2$ &  & $K_1$ & $K_2K_3$ & & $\Gamma M_1$ &$\Gamma M_2 $ \tabularnewline
\hline 
$E$ & 1 & 1 & 2 &        $E$ & 1 & 1 &        $E$ & 1 & 2  &        $E$  & 1&1    \tabularnewline
$2C_{3}$ & 1 & 1 & -1 &  $C_{2}$ & 1 & -1 & $C_{3}$ & 1 & -1 &  $C_{2}$& 1&-1\tabularnewline
$3C_{2}$ & 1 & -1 & 0 &     &   &   &       $C_{3}^{-1}$ & 1 & -1 & & & \tabularnewline
\hline
\end{tabular}
\protect\caption{\label{tab:irreps-MSG} Character table of irreducible representations at high symmetry momenta and lines in magnetic space group $P6^\prime2^\prime2$. $E$, $C_3$, and $C_2$ represent the conjugation classes generated from identity, $\mathcal{C}_{3z}$ and $\mathcal{C}_{2x}$. The notation $\Gamma M_i$ stands for the symmetric line $\Gamma-M$ . }  
\end{table}

Based on the generators discussed so far, the symmetry group of the moir\'e lattice is the magnetic space group called $P6^{\prime}2^{\prime}2$ (\#177.151 in the BNS notation~\cite{Bilbao-MSG}). Although the same group describes moir\'e bilayer, the physics is different here. It indeed originates from $3$ Dirac cones, instead of two, and the extra particle-hole symmetry is essentially different. The high-symmetry points and their little co-groups are 
$\Gamma$ ($\mathcal{C}_{2x}, \mathcal{C}_{3z}, \mathcal{C}_{2z} \mathcal{T}, \mathcal{P}$), $K_M$ ($\mathcal{C}_{3z}, \mathcal{C}_{2z} \mathcal{T}, \mathcal{P}\mathcal{C}_{2x}$) and $M$ ($\mathcal{C}_{2x},  \mathcal{C}_{2z} \mathcal{T}, \mathcal{P}$).
The symmetries on the high-symmetry lines are $\Gamma - M$ ($\mathcal{C}_{2x}$, $\mathcal{C}_{2z} \mathcal{T}$) and
$\Gamma - K_M$ ($\mathcal{C}_{2x} \mathcal{P},  \mathcal{C}_{2z} \mathcal{T}$).
The classification of the different irreducible representations at the symmetric points and lines are given in Table~\ref{tab:irreps-MSG}. At $\Gamma$, $M$, $K_M$ and on the line $\Gamma-M$, each energy or band in  Fig.~\ref{fig2} is characterized by  a certain representation determined from the character, {\it i.e.} from the eigenvalues of the operators $\mathcal{C}_{3z}$ and $\mathcal{C}_{2x}$ restricted to this (possibly degenerate) energy.

\noindent {\it First proof of all-connected bands.} We now prove by contradiction that all bands are connected such that there is no gap in the spectrum at any energy. We assume a subspace of isolated bands between the energies $\varepsilon_{1,k}$ and  $\varepsilon_{2,k}$. By p-h symmetry, a symmetric set of bands exists in the energy window $(-\varepsilon_{2,-k},-\varepsilon_{1,-k})$ and, consequently, the $N_1$ bands between $-\varepsilon_{1,-k}$ and  $\varepsilon_{1,k}$ must be disconnected from all other bands. We focus on these $N_1$ bands and investigate their transformation property under $\mathcal{C}_{2x}$. $\mathcal{C}_{2x}$ remains a symmetry along the line connecting $\Gamma$ and $M$ such that their total character must coincide at each end, or $\chi_{\mathcal{C}_2\Gamma}=\chi_{\mathcal{C}_2M}$.

We call $m_{\Gamma_i}$ the multiplicity of the representation $i=1,2,3$ in the set of $N_1$ bands, and $m_{M_i}$ the multiplicity of the representation $i=1,2$ in the set of $N_1$ bands at $M$. Hence we have the equations:
$ m_{\Gamma_1} + m_{\Gamma_2} + 2 m_{\Gamma_3}  = N_1$, $m_{M_1}+ m_{M_2} = N_1$, $m_{\Gamma_1} - m_{\Gamma_2}= \chi_{\mathcal{C}_2\Gamma}$ and $m_{M_1}- m_{M_2} =\chi_{\mathcal{C}_2M}$
which leads to
\begin{equation}\label{eq:absurdum}
  m_{M_2} =  m_{\Gamma_2} +  m_{\Gamma_3}.
\end{equation}
For $\alpha = 0$, we have three Dirac cones around $\Gamma$, $K_M$ and $K_M'$ and a gapped spectrum at $M$. At $\Gamma$, the zero energy subspace is doubly degenerate and the character of  $\mathcal{C}_{3z}$ is simply given by ${\rm Tr}  ( e^{i \frac{2\pi}{3} \sigma_z} ) = -1$
corresponding to the irreducible representation $\Gamma_3$ as indicated in Table~\ref{tab:irreps-MSG}. We thus have $m_{\Gamma_3} = 1$, $m_{M_2} =  m_{\Gamma_2}=0$ when restricted to zero energy and $\alpha = 0$. Increasing $\alpha$ away from zero, the $\mathcal{P}$ symmetry is maintained and commutes with $\mathcal{C}_{3z}$ and $\mathcal{C}_{2x}$ such that any band associated to a given representation collapsing at (or departing from) zero energy, at either $\Gamma$ or $M$, must move with its energy symmetric p-h partner associated to the same representation. As a result, the multiplicities $m_{\Gamma_j}$ and  $m_{M_j}$ can only change by units of two in such processes.
The same argument extends to non-zero energies where each band with energy $\varepsilon$ and representation $\Gamma_j$ (or $M_j$) has a p-h partner with energy $-\varepsilon$ and the same representation. Since $k$ and $-k$ are identified at $\Gamma$ and $M$ such that the interval $(-\varepsilon_{1,-k},+\varepsilon_{1,k})$ is symmetric, we finally obtain by continuity with $\alpha$  that the multiplicity  $m_{\Gamma_3}$ must be odd while $m_{M_2}$ and $m_{\Gamma_2}$ are both even integers. It contradicts Eq.~\eqref{eq:absurdum}, thus completing the proof.

\noindent {\it General proof for unequal twist angles.} Our discussion was so far restricted to the symmetric configuration of equal $p$ and $q$. The $\mathcal{C}_{2x}$ and $\mathcal{P}$ symmetries are broken when $p$ and $q$ are different whereas $\mathcal{C}_{3z}$ and $\mathcal{C}_{2z} \mathcal{T}$ are maintained. However, a remnant of p-h symmetry still exists in the form of a mirror symmetry $\Pi_{\pi/6}$ with respect to the plane orthogonal to the layer and crossing $\Gamma$ and $K_M$, leaving each layer invariant. It corresponds to the operator $\bm{\Pi}_{\pi/6} = \mathcal{P} \mathcal{C}_{2x} \mathcal{C}_{3z}$ with $\bm{\Pi}_{\pi/6}^2=1$, acting as 
\begin{equation}
\bm{\Pi}_{\pi/6} \tilde H(k) \bm{\Pi}_{\pi/6}^\dagger = -  \tilde H(\Pi_{\pi/6} k),
\end{equation}
which associates pairs of mirror-symmetric momenta with opposite energies. Extending the arguments of Ref.~\cite{ahn2018}, we show that a set of isolated bands with $C_{2z} T$ symmetry cannot accommodate an odd winding number $N_t$, corresponding for example to an odd number of Dirac cones. The derivation is explicit in Ref.~\cite{song2018,ahn2018} for two bands and a vanishing total Berry phase (Wilson loop) with $N_t = - 2 e_2$, where the Euler class $e_2$ is an integer topological invariant. Adding more bands, windings around singularities can change sign but keep a definite parity while the parity of $e_2$ defines a $\mathbb{Z}_2$ topological invariant, the Stiefel-Whitney class $w_2$. Then, the relation $N_t = - 2 w_2$ simply enforces that the winding number must be an even integer. More intuitively, we note that Dirac points are monopoles attaching Dirac strings. They can annihilate in pairs when of opposite signs or form a topological isolated band by combining pairs of same sign~\cite{song2018,Po2018b}, but in all cases they need to pair to form an isolated set of bands. 

We now show by contradiction that all bands are connected by gapless points in our trilayer moir\'e model for arbitrary $p$ and $q$. As already discussed above, we can assume, without loss of generality, a set of disconnected bands symmetric around zero energy. P-h symmetry implies that all band crossings at non-zero energy come in pair such that the analysis of the parity of $N_t$ can be restricted to zero-energy modes. A single Dirac cone is protected by $\mathcal{C}_{2z} \mathcal{T}$ and is pinned at zero energy by p-h sP. p-h further protects the parity of $N_t$ for zero modes as $\alpha$ is varied. By continuity with the case $\alpha=0$ where we have three Dirac cones and $N_t=3$, we finally obtain that $N_t$ is odd, in contradiction with  $N_t = - 2 w_2$, which completes our proof that all bands must be connected.


In summary, we showed that trilayer twisted graphene exhibits band flattening along symmetry lines and close to magic angles. We also proved, by compatibility of band representations for evenly twisted planes or by counting an odd number of Dirac points protected by $\mathcal{C}_{2z} \mathcal{T}$, that the system is always a metal with an infinite connectivity, an unprecedented feature in standard materials~\cite{Bernevig_TQC,vergniory2018,zhang2018,tang2018}. This property relies on p-h symmetry emerging for small twisting angles. We checked that this condition is practically realized already for angles close to the first magic angle~\footnote{See Supplemental Material with the numerical investigation of p-h symmetry breaking, more details on the symmetries and the positions of the Dirac points at $\alpha=0$.}. Since it originates from the underlying three Dirac cones, we conjecture that the property of infinite band connectivity will appear in many other configuration such as multilayer moir\'e graphene with an odd number of planes.

We would like to thank B. Estienne and Biao Lian for fruitful discussions. BB is supported by the Department of Energy Grant No. de- sc0016239, the National Science Foundation EAGER Grant No. noaawd1004957, Simons Investigator Grants No. ONRN00014-14-1-0330, No. ARO MURI W911NF- 12-1-0461, and No. NSF-MRSEC DMR- 1420541, the Packard Foundation, the Schmidt Fund for Innovative Research.

\bibliography{ref}

\onecolumngrid

\newpage

\begin{center}
{\bf Supplementary information for: Flat bands and perfect metal in trilayer moir\'e graphene}
\end{center}

\renewcommand{\thesection}{S-\Roman{section}}
\renewcommand{\theequation}{S-\arabic{equation}}
\renewcommand{\thefigure}{S-\arabic{figure}}

\section{Breaking particle-hole symmetry}

The particle-hole (p-h) symmetry is important in proving that all bands in the spectrum of trilayer twisted graphene are connected. It gives a band landscape symmetric with respect to zero energy and implies that features, such as Dirac points and other gapless crossings, or symmetries occurring at finite energy always come in pairs. It maintains globally the gapleass structure of the spectrum. The other essential symmetry is the combined spatial-time symmetry $\mathcal{C}_{2z} \mathcal{T}$ - this symmetry locally maintains the gapless structure of the band spectrum. Breaking $\mathcal{C}_{2z} \mathcal{T}$ directly open gaps at Dirac points locally and hence isolates finite sets of bands. In contrast to that, we expect the gapless band structure to keep its integrity if the terms breaking p-h symmetry are not too strong.

\begin{figure}
\begin{centering}
\includegraphics[width=.9\linewidth]{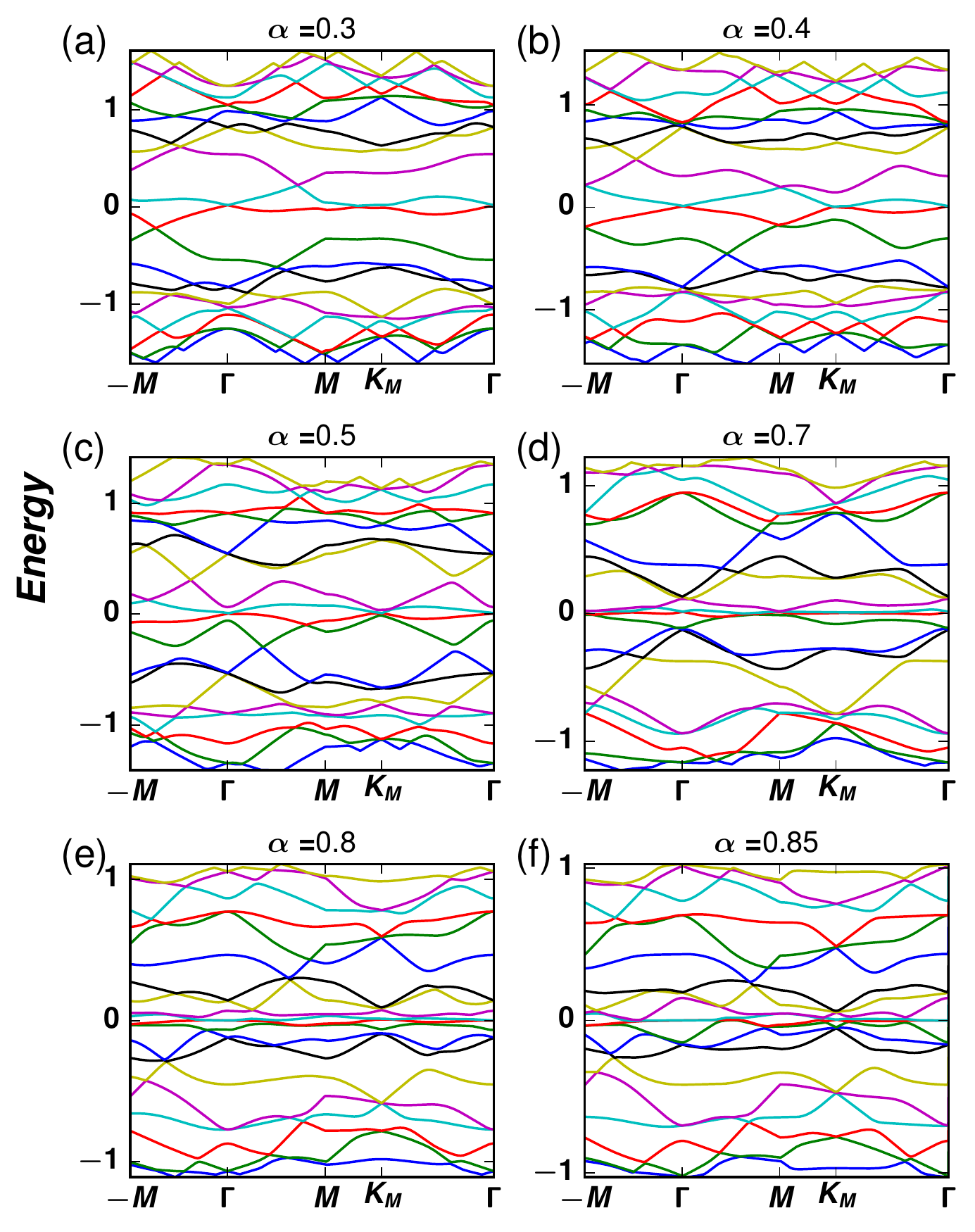}
\par\end{centering}
\caption{Moir\'e bands computed from Eq.~\eqref{hamiltonianbetweenlayesrab2} for $p=q=1$ and $\theta_{1,2} = \theta_{23}$. $20$ bands around zero energy are represented along the moir\'e Brillouin zone trajectory $-M \to \Gamma \to M \to K_M \to \Gamma$ for $\alpha = 0.3, 0.4, 0.5, 0.7, 0.8, 0.85$ (a-f). In principle, Eq.~\eqref{hamiltonianbetweenlayesrab2} has two independent parameters, $\alpha$ and $\theta_{12}$, and the ratio between them is not universal. For this plot, we have nevertheless chosen the conversion rule $\theta_{12} = 1.05\degree$ for $\alpha=0.606$ compatible with experiment results on graphene structures. \label{fig1-s}}
\end{figure}

In a real trilayer graphene system, the rotation angles between the planes are finite and there are corrections to the emergent Eq.~\eqref{hamiltonianbetweenlayesrab} of the main text. There are two main effects that break p-h symmetry: (i) $k^2$ corrections to the Dirac cones approximation $\sim  \mathbf{k}\cdot\boldsymbol{\sigma}$ close to the $K$ points in each layer, (ii) the three K points are not aligned at finite rotation angle. We note that (i) is the first step towards the complete band structure of each graphene layer with both the K and K' points. The vicinities of K and K' are indeed fully decoupled in Eq.~\eqref{hamiltonianbetweenlayesrab} of the main text. (ii) can be simply taken into account by choosing a different basis in each layer such that the vectors ${\bf q}_1^{ab}$ are parallel. This results into the Hamiltonian
\begin{equation}
  H^{(ab)}(\delta \mathbf{p}_a, \delta \mathbf{p}_b)  = v_{F}  \left(M_{\theta_{1a}}^{-1}\delta\mathbf{p}\right) \cdot\boldsymbol{\sigma}\
  \delta_{a,b}  +  w^{ab} \sum_{j=1}^{3}\delta_{\delta\mathbf{p}_a,\delta\mathbf{p}_b+\mathbf{q}_{j}^{a,b}}T^{j} \label{hamiltonianbetweenlayesrab2}
\end{equation}
where $M_{\theta_{1a}}$ is the in-plane rotation of angle $\theta_{1a}$, reducing to Eq.~\eqref{hamiltonianbetweenlayesrab} of the main text for vanishingly small twist angles. Eq.~\eqref{hamiltonianbetweenlayesrab2} explicitely breaks p-h symmetry. The numerical resolution of Eq.~\eqref{hamiltonianbetweenlayesrab2} is displayed in Fig.~\ref{fig1-s} for the representative case $p=q=1$ and for several values of $\alpha$. Inspecting the $20$ states close to the Fermi energy along symmetry lines, we find band crossings between all states despite the breaking of p-h symmetry.

\section{Symmetries for arbitrary $p$ and $q$}

We first review the particular case $p=q=1$.
The effective single-valley model of Eq.~\eqref{hamiltonianbetweenlayesrab} (see main text) is obtained by keeping only the states close to the K point in each layer.
It is decoupled from its time-reversed counterpart built with the states around the K' points.
Hence, Eq.~\eqref{hamiltonianbetweenlayesrab} does not respect time-reversal symmetry.
For $p=q=1$, the symmetry group of this one-valley model is the magnetic space group $P6^{\prime}2^{\prime}2$, the same as for twisted bilayer graphene, characterized by the generators 
\begin{equation}\label{symgen}
\mathcal{C}_{3z} =\exp(i 2\pi\sigma_z /3 ) \delta_{Q_m, C_{3z} Q_n}, \qquad \mathcal{C}_{2x} =\sigma_x \delta_{Q_m, C_{2x} Q_n}, \qquad \mathcal{C}_{2z} \mathcal{T} =\sigma_x \delta_{Q_m, Q_n} K.
\end{equation}
The band spectrum is invariant with respect to these three symmetries. The antiunitary $C_{2z} T$ symmetry is a combination of time-reversal and $C_{2z}$ symmetry acting locally on the moir\'e lattice. It commutes with the spatial symmetries $\mathcal{C}_{3z}$ and $\mathcal{C}_{2x}$. Whereas $\mathcal{C}_{3z}$ maps each layer into itself, $\mathcal{C}_{2x}$ exchanges the layers $1$ and $3$ and leaves the layer $2$ invariant.

Besides the magnetic space group $P6^{\prime}2^{\prime}2$, the model also exhibits a particle-hole (p-h) symmetry inherited from the original p-h symmetry in each graphene layer. The corresponding operator - see Eq.~\eqref{phsym} and \eqref{phsym2} in the main text -
\begin{equation}
\mathcal{P} = \delta_{Q_m, -Q_n}\zeta_{Q_n}, \qquad  \qquad \mathcal{P}   \tilde H(k) \mathcal{P}^\dagger = -  \tilde H(-k),
\end{equation}
commutes with the generators \eqref{symgen} of the magnetic space group, and squares to $1$ in contrast with the p-h operator in the bilayer case which squares to $-1$.

The rotation symmetry $\mathcal{C}_{3z}$ and the antiunitary symmetry $C_{2z} T$ are valid also for arbitrary $p$ and $q$, {\it i.e.} when the rotation angles are different but still commensurate. Interestingly, if $p$ or $q$ is different from $1$, the symmetries $\mathcal{C}_{2x}$ and $\mathcal{P}$ are individually broken but their product $\mathcal{C}_{2x} \mathcal{P}$ remains a symmetry of the one-valley Hamiltonian~\eqref{hamiltonianbetweenlayers} (see main text).  We thus introduce the mirror p-h symmetry operator
\begin{equation}
\bm{\Pi}_{\pi/6} = \mathcal{P} \mathcal{C}_{2x} \mathcal{C}_{3z} = \begin{pmatrix} 0 & e^{-2 i \pi/3} \\ e^{2 i \pi/3} & 0 
\end{pmatrix} \delta_{Q_m,  \Pi_{\pi/6} Q_n} \zeta_{Q_n} ;\;\;\;\ \zeta_{Q_1}= \zeta_{Q_3} = 1;\;\;\; \zeta_{Q_2}=-1,
\end{equation}
obtained by combining the  product $\mathcal{C}_{2x} \mathcal{P}$ with the rotation symmetry $\mathcal{C}_{3z}$. $\zeta_{Q_n}$ is $+1$ for $Q_n$ belonging to the bottom and top layers and $-1$ for the middle layer.  The invariant plane of the mirror symmetry $\Pi_{\pi/6}$ crosses orthogonally the graphene layers along lines. These lines make an angle of $\pi/6$ with respect to the $x$ axis and go through the $\Gamma$ and $K_M$ points. By construction, each layer is thus invariant upon the mirror symmetry $\Pi_{\pi/6}$. The action  on the Hamiltonian~\eqref{hamiltonianbetweenlayers} is
\begin{equation}\label{phsym2-2}
\bm{\Pi}_{\pi/6} \tilde H(k) \bm{\Pi}_{\pi/6}^\dagger = -  \tilde H(\Pi_{\pi/6} k),
\end{equation}
corresponding to a p-h symmetry. Eq.~\eqref{phsym2-2} implies a symmetric band spectrum around zero energy where each state $k$ has a mirror-symmetric partner $\Pi_{\pi/6} k$ with opposite energy.

\section{Positions of the Dirac cones for $\alpha=0$}

\begin{figure}
\begin{centering}
\includegraphics[width=0.95\linewidth]{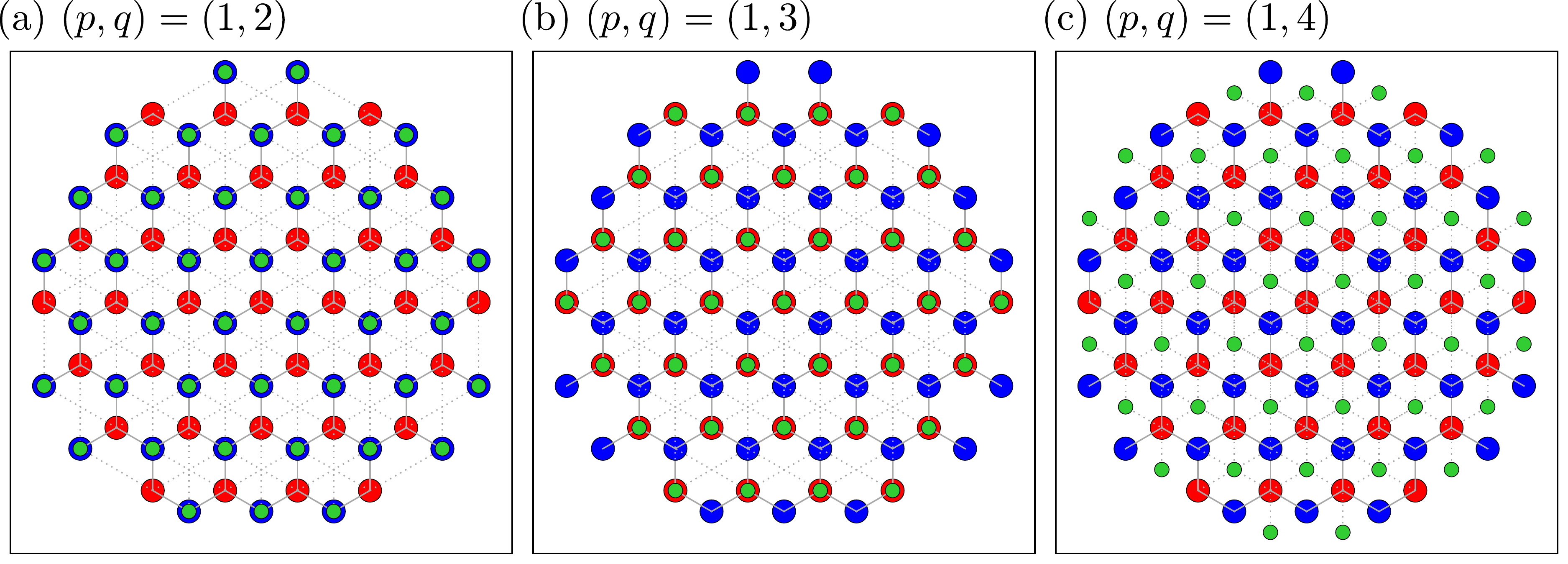}
\par\end{centering}
\caption{\label{fig_lattices} Three lattices of Dirac cones for $\alpha=0$ corresponding to different  $(p,q) = (1,2)$ (a), $(1,3)$ (b), and $(1,4)$ (c). Red (resp. blue, green) points correspond to the middle (resp. bottom, top) layer. The right lattice (c) has separate Dirac cones, corresponding to the case (i) (see supplementary main text), whereas the left (a) and middle (b) lattices have pairs of overlapping Dirac cones and $1$ isolated Dirac cone, corresponding to the case (ii) of the supplementary main text.}
\end{figure}

As discussed in the main text, the full connectivity of the moir\'e band model requires essentially the $C_{2z} T$ symmetry to protect Dirac cones, or at least to have them gapped by pairs, the mirror p-h symmetry $\bm{\Pi}_{\pi/6}$ to reduce the parity of Dirac points to their parity in the zero energy manifold, and the continuity with respect to the case $\alpha=0$ where the positions of Dirac points are easily determined. As illustrated in Fig.~\ref{fig_lattices}, we find two configurations at $\alpha=0$: either (i) the model has three Dirac points at distinct positions $\Gamma, K_M, K_M'$ in the Brillouin zone, or (ii) two Dirac points are on top of each other and separated from the third one. The case (i) occurs for $q-p = 1 \mod 3$ whereas (ii) occurs for $q-p = 0,2 \mod 3$.

\section{Densities of electrons close to the Fermi energy}

\begin{figure}
\begin{centering}
\includegraphics[width=0.7\linewidth]{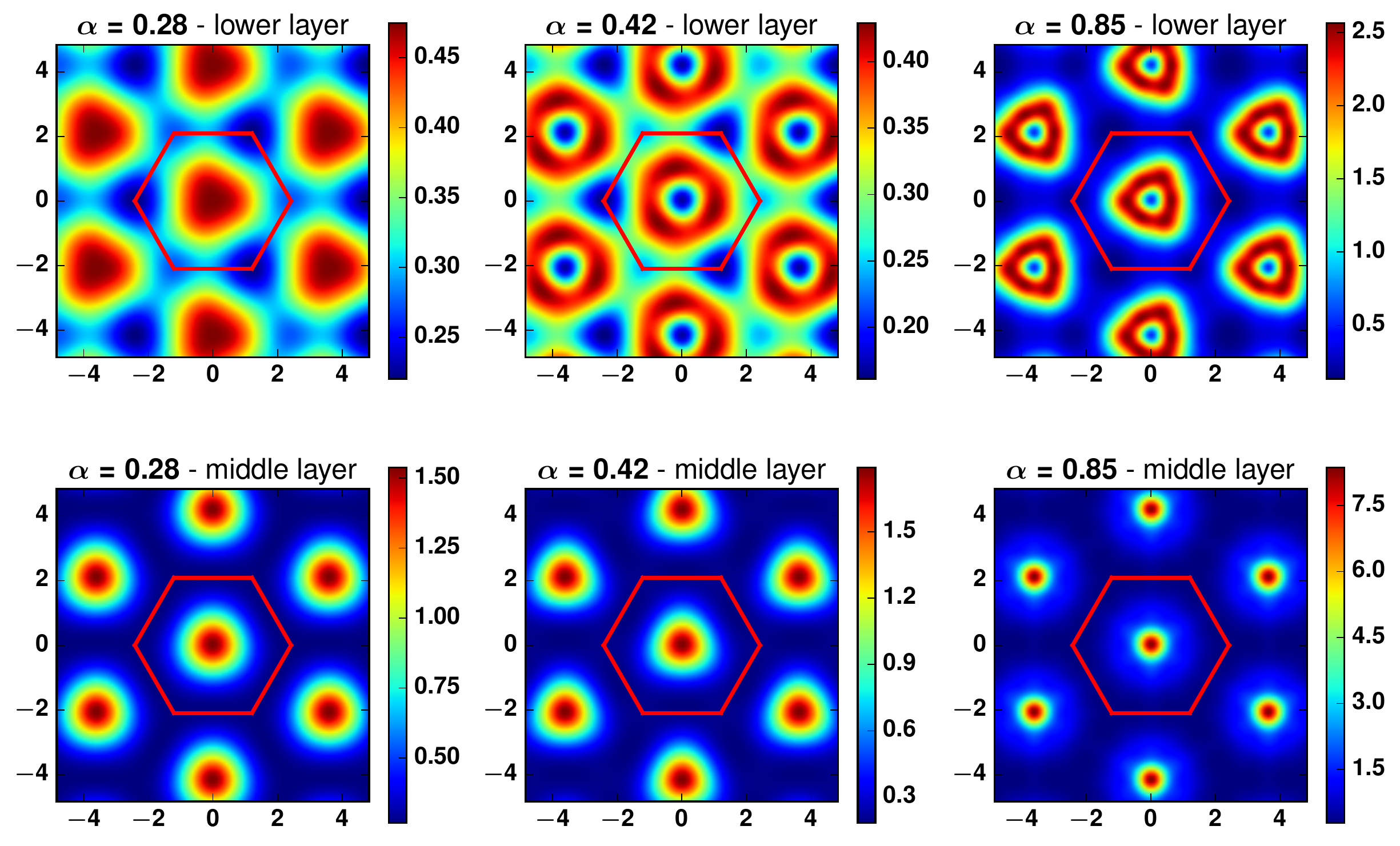}
\par\end{centering}
\caption{\label{fig_LDOS} Local density of states for the $A$ sublattice, and $p=q=1$, integrated for $|\varepsilon| < \varepsilon_{\rm max}$. Its counterpart for the $B$ sublattice is infered by $C_6$ symmetry, and the upper layer one is obtained from the lower layer by particle-hole symmetry. We choose $\varepsilon_{\rm max} = 0.02$ and $\alpha = 0.28, 0.42, 0.85$.}
\end{figure}

Twisted bilayer graphene, with a small angle, exhibits a local density of states for the (almost) flat bands that is well-localized to the AA regions of the Moir\'e pattern~\cite{laissardiere2010,TB_TBG,cao2016}, forming a triangular lattice. This spatial localization is understood at zero energy from the absence of tunneling between AA and AB, BA regions~\cite{dossantos2012}. In the trilayer geometry, it is not possible to isolate a finite set of bands as we prove in the main text. However, we can integrate the local density of states (LDOS) over a narrow energy range close to zero energy similarily to what has been done in the bilayer case~\cite{dossantos2012}. The result is shown in Fig.~\ref{fig_LDOS} for $\alpha = 0.28, 0.42, 0.85$. In the middle layer, localization around the $AA$ region occurs more dramatically when bands are nearly dispersionless close, see $\alpha = 0.85$. Although the whole pattern forms a triangular lattice, the localization around the $AA$ region is never really strong for the lower and top layers.

\end{document}